# Engineered creation of periodic giant, non-uniform strains in MoS$_2$ monolayers

Elena Blundo[†], Cinzia Di Giorgio[‡,§], Giorgio Pettinari[#], Tanju Yildirim[◊], Marco Felici[†], Yuerui Lu[◊], Fabrizio Bobba[‡,§] and Antonio Polimeni[†,*]

[†]Physics Department, Sapienza University of Rome, 00185 Rome, Italy.
E-mail: antonio.polimeni@roma1.infn.it

[‡]Department of Physics E.R: Caianiello, University of Salerno, 84084 Fisciano, Italy.

[§] SuPerconducting and other INnovative materials and devices institute, National Research Council, 84084 Fisciano, Italy.

[#]Institute for Photonics and Nanotechnologies, National Research Council, 00156 Rome, Italy.

[◊]Research School of Electrical, Energy and Materials Engineering, College of Engineering and Computer Science, The Australian National University, Canberra, ACT2601, Australia.

## Abstract

The realization of ordered strain fields in two-dimensional crystals is an intriguing perspective in many respects, including the instauration of novel transport regimes and the achievement of enhanced device performances. In this work, we demonstrate the possibility to subject micrometric regions of atomically-thin molybdenum disulphide (MoS$_2$) to giant strains with the desired ordering. Mechanically-deformed MoS$_2$ membranes can be obtained by proton-irradiation of bulk flakes, leading to the formation of monolayer domes containing pressurized hydrogen. By pre-patterning the flakes via deposition of polymeric masks and electron beam lithography, we show that it is possible not only to control the size and position of the domes, but also to create a mechanical constraint. Atomic force microscopy measurements reveal that this constraint alters remarkably the morphology of the domes, otherwise subject to universal scaling laws. Upon the optimization of the irradiation and patterning processes, unprecedented periodic configurations of large strain gradients -estimated by numerical simulations- are created, with the highest strains being close to the rupture critical values (> 10 %). The creation of such high strains is confirmed by Raman experiments. The method proposed here represents an important step towards the strain engineering of two-dimensional crystals.

KEYWORDS: strain, MoS$_2$, two-dimensional materials, engineering, domes, patterning.

**Introduction.**

In a solid-state environment, the possibility to engender novel fundamental effects or to achieve favourable conditions for the realization of devices typically relies on the ability to modify the basic properties of crystalline materials. The creation of alloys and heterostructures, the controlled introduction of dopants, and strain are common tools. Among them, strain is particularly relevant for two dimensional (2D) crystals -such as graphene and transition metal dichalcogenides (TMDs)- where flexibility and robustness[1,2] couple to a high sensitivity to mechanical deformations[3]. For TMDs, which are direct gap semiconductors[4,5,6] characterized by a strong spin-orbit coupling[7] in the monolayer (ML) limit, tensile strain leads to a sizeable reduction and modification of the band gap[8,9,10,11,12], up to turning it from direct to indirect[13,14,15,16]. Extremely high strains[2] ($\gtrsim 10$ %) can be applied in a reversible manner, and absorption, emission and carriers' mobility and lifetime can thus be modulated[10,13,17,18] on-demand. Therefore, the application of strain-engineering protocols to TMDs holds a great potentiality for optoelectronics, spin- and valley-tronics. Several methods for straining atomically-thin-materials have been developed, including bending apparatuses[11,19], deposition on pillars[20], epitaxial growth of superlattices[21], and creation of bulges by means of specific devices[10,22], or spontaneous formation of domes[23,24,25]. Bulges or domes exploit the pressure exerted on the crystal by trapped gases and nowadays represent the most efficient method for achieving highly strained 2D membranes, also allowing a certain degree of control over the spatial distribution of the deformation[10,24]. This is particularly relevant for ML-TMDs, where the creation of strain gradients leads to a seamless reduction of the band gap[13,24] and can result in pseudo-gauge fields able to rule the quasiparticles' motion[26]. In turn, stable and periodic strain modulations are promising for broadband light absorption and carriers' harvesting[13,18], and for the generation of persistent currents[27], which make them on the one side apt to photovoltaics, photocatalysis and photodetection devices[18], on the other side a fruitful platform for the observation of novel physical phenomena. However, the great promise of these systems is hindered by the fact that the durability of the bulges is limited to a few weeks[10,22], whilst the deformation of the domes is governed by energy minimisation[23,28], that establishes an upper bound for the achievable strains. Here, we move a step further by developing a method to achieve unprecedented periodic, durable and extremely high strain gradients in $MoS_2$ through the engineered creation of domes.

**Results and discussion.**

In this paper, bulk flakes are mechanically exfoliated on $SiO_2$/Si substrates and proton-irradiated -as in Ref. 24- resulting in the formation of isolated spherical domes or conglomerates of domes on the flake surface, as shown in Figure 1a. The domes are thick just one S-Mo-S layer and their size varies from few tens of nanometres to several micrometres[24]. In Ref. 24, the use of polymeric masks allowed the creation of ordered arrays of $WS_2$ domes with control over their size and position. Here, several samples were prepared by depositing hydrogen silsesquioxane (HSQ) negative-tone e-beam resists with different thicknesses -in between 20 and 100 nm- on $MoS_2$ bulk flakes. Octagonal openings with different sizes were then created in the masks via electron-beam lithography (EBL). The choice of the resist and the patterning procedure are discussed in Supplementary Information, Note 1. The samples were then proton-irradiated. As shown in Figures 1b-d for an 80-nm-thick resist with openings of size $S = 5$ μm, 3 μm, and 1 μm, respectively, ordered arrays of isolated $MoS_2$ domes with different sizes can be created during the same irradiation process. As shown in Supplementary Fig. S1, second harmonic generation measurements were used to verify that the patterned domes are just 1-layer-thick[29,30].

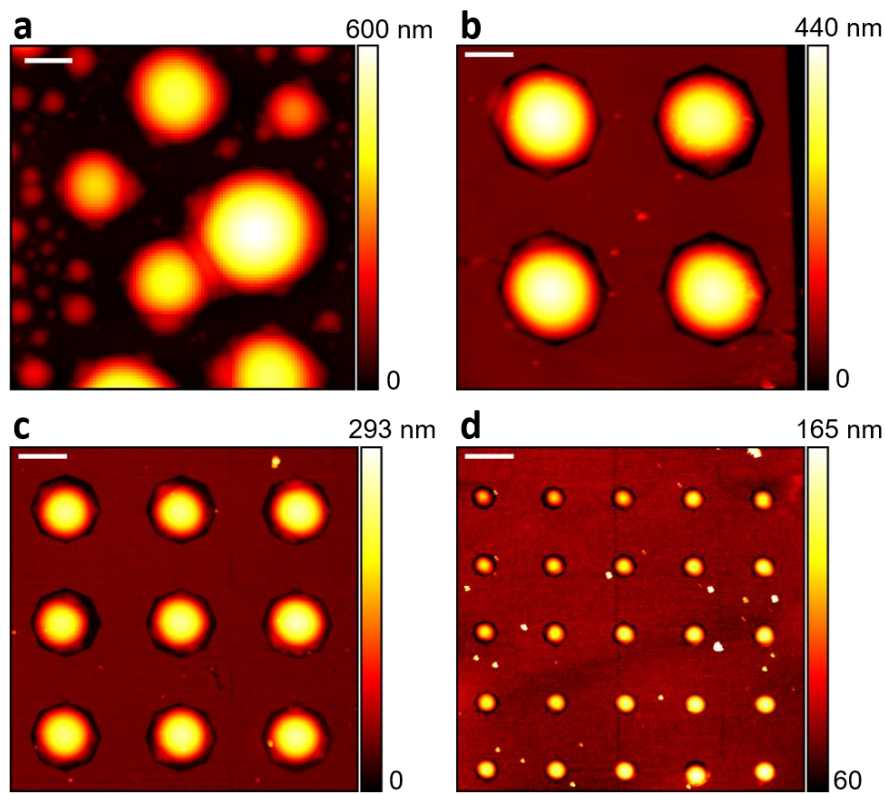

**Figure 1.** Formation of random and patterned $MoS_2$ domes. (a) AFM image of a 15x15 μm² area of a proton-irradiated $MoS_2$ flake. (b-d) Same for $MoS_2$ flakes patterned with an 80-nm-thick mask with openings of size $S = 5$ μm (b), 3 μm (c), and 1 μm (d). The scalebars correspond to 2 μm, while the origin of the z axis is set at the flake surface.

Differently from Ref. 24, where the domes created in the patterns had much smaller dimensions than the opening size, the domes in Figures 1b-d are almost completely filling the openings. This result has been achieved by optimizing the proton-dose, as shown in Figure 2 for a sample patterned with a 30-nm-thick resist and having openings with size $S = 5$ μm, 3 μm, and 1 μm. First, the sample was irradiated with dose $d_0 = 5.5 \cdot 10^{16}$ protons/cm² (higher than in Ref. 24), as schematized in Fig. 2a, resulting in the formation of domes still not filling the openings, (see AFM image shown as inset).

The sample was then re-irradiated with dose $d_1 = 2.0 \cdot 10^{16}$ protons/cm$^2$, and the domes' footprint was brought to nearly totally occupy the openings (see Fig. 2b). A comparison between the filling percentages after the first and second irradiation is shown in Fig. 2c, while a comparison with the sample of Ref. 24 is shown in Supplementary Figure S2. Following this result, all the samples were irradiated with doses equal to 6-7·10$^{16}$ protons/cm$^2$, representing the best trade-off between high filling and high formation yield (for too high doses, the domes start exploding, see Supplementary Note 1). The morphological properties of the domes were probed by atomic force microscopy (AFM, see Experimental Section). As theoretically predicted[23,28] and experimentally demonstrated,[23,24,25] atomically thin domes feature a universal aspect ratio ($h_m/R$, with $h_m$ maximum height and $R$ footprint radius) which remains constant independently of the radius value. For MoS$_2$ random domes on MoS$_2$ flakes, the universal $h_m/R$ value is 0.16[23,24]. We find the same average value for random domes (as exemplified by the dome whose profile is shown in Figure 3a, with $h_m/R = 0.165$) and for the patterned domes of Fig. 2, even when filling the openings, implying that a 30-nm-thick resist does not alter the mechanics of the formation process. Samples with thicker resists were then prepared: The three patterned domes in Figs. 3b-d were created in masks thicker than 50 nm, and have $h_m/R = 0.186$ ($S = 5$ μm, Fig. 3b), 0.219 ($S = 3$ μm, Fig. 3c), 0.264 ($S = 1$ μm, Fig. 3d). Interestingly, in this case the mask acts as a strong constraint over the spontaneous formation of the domes and allows to overcome the natural, universal limit.

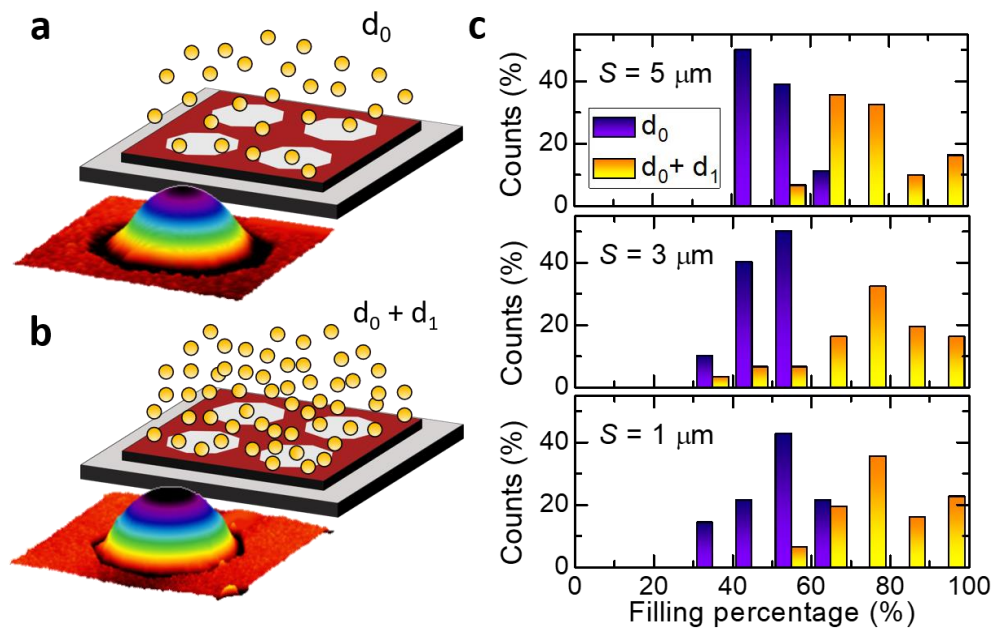

**Figure 2.** Optimization of the proton-irradiation process. (a) Sketch of a patterned (wine area) sample (gray part) irradiated with protons (orange particles) with dose $d_0$. The inset is an AFM 3D image of a typical dome (in an opening with diameter $S = 5$ μm) formed in a sample patterned with a 30-nm-thick mask and irradiated with proton dose $d_0 = 5.5 \cdot 10^{16}$ protons/cm$^2$. The dome is not totally filling the opening, as demonstrated by the presence of the black region around the dome footprint. (b) Same as panel a, but for the use of a higher proton dose $d_0+d_1$, with $d_1 = 2.0 \cdot 10^{16}$ protons/cm$^2$. In this case, a typical dome (in an opening with diameter $S = 5$ μm) almost entirely occupies the opening. (c) Histograms of the filling

percentages (footprint of the dome divided by the area of the opening) for the sample described in panel a after irradiation with proton dose $d_0$ (blue columns) and $d_0+d_1$ (orange columns), for different opening sizes.

This is particularly relevant because the higher the deformation, the higher the strain values and strain gradients. More in detail, atomically thin domes are characterized by an anisotropic tensile in-plane strain that increases from the edge towards the summit[13,24,31]. At the summit, strain becomes isotropic biaxial and proportional to the square of the aspect ratio, according to both the model developed by Hencky[10,32] and numerical simulations[24,31]. Due to the analogous effect of the different in-plane strain components (*i.e.*, the radial, $\varepsilon_r$, and circumferential, $\varepsilon_t$, components[24]) on the electronic properties of TMDs[8], one can quantify the strain acting on the 2D membrane by translating the anisotropic strain tensor into a total in-plane strain defined as[13] $\varepsilon_p = \varepsilon_r + \varepsilon_t$.

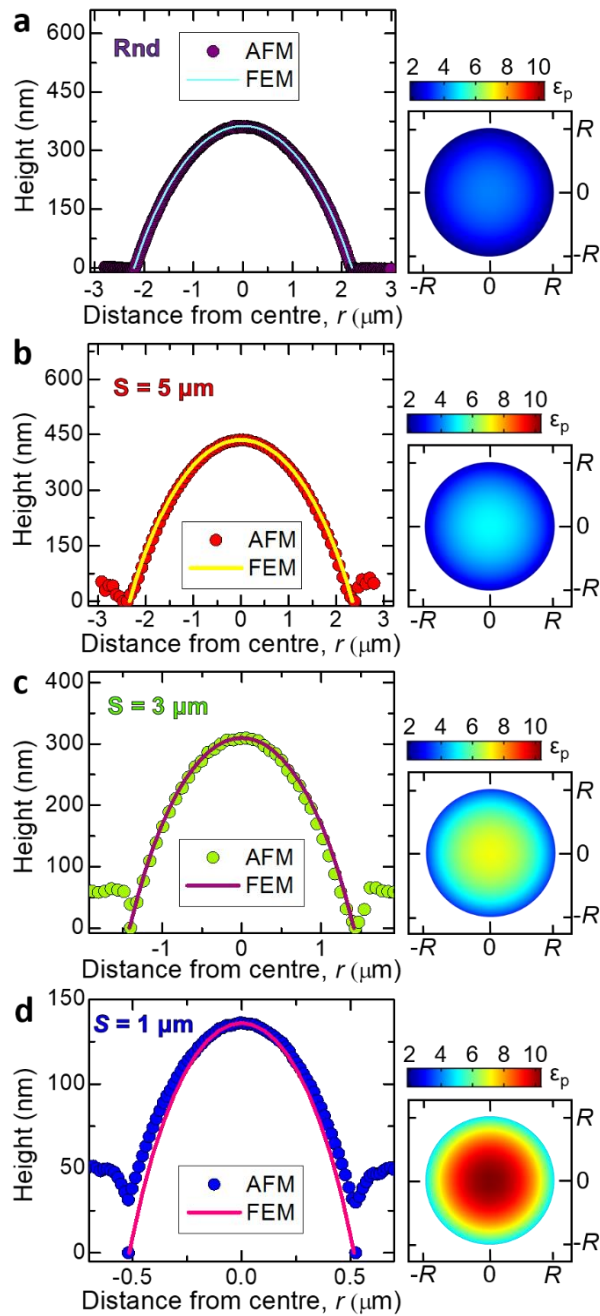

**Figure 3.** Height profile and total in-plane strain in MoS$_2$ domes. (a-d) Left panels: AFM profiles (coloured dots) along a diameter of a random (a), and three patterned domes created in openings with size $S = 5$ µm (b), $S = 3$ µm (c), and $S = 1$ µm (d). For the dome in panel d the AFM tip was not able to measure its base and the two points at 0 were set by measuring the thickness of the resist in a nearby corridor. The ratio between the vertical and horizontal axis is the same for the 4 plots, to emphasize the differences in the aspect ratios of the domes. The solid lines represent the profiles calculated via FEM simulations. Right panels: spatial distribution of the total percentage in-plane strain for the domes on the left, according to the colour bar displayed on top. The same scale has been used for the 4 plots.

According to Hencky's model, MoS$_2$ domes are characterized at the summit by a total in-plane strain[24] $\varepsilon_{\mathrm{p}} = 2 \cdot f(\nu) \cdot (h_{\mathrm{m}}/R)^2$, with $f(\nu) = 0.721$, while numerical methods have to be employed to calculate the strain field distribution across the domes' surface. For this reason, finite-element method (FEM) calculations[24] were used to simulate the profiles of the domes (that are in good agreement with the AFM profiles, see Fig. 3) and calculate the strain components, as detailed in Refs. 13 and 24. In the right panels of Figs. 3a-d, we show the spatial distribution of the total in-plane strain associated to the domes in the left panels, while calculations concerning the geometry and degree of isotropy of the strain tensor and its effect on the exciton transitions are discussed in Supplementary Note 2. Our calculations show how extremely high strains -larger than 10 %- are achieved for domes created in small openings ($S = 1$ µm). To our knowledge, such high strains - close to the rupture limit- could only be achieved so far by indentation of suspended monolayers[2] or with bulging devices[10], but in this latter case the structures are not durable[10,22]. With our method, instead, the domes typically last for years[24] (part of the measurements performed on a sample in this work was taken two years after creating the domes).

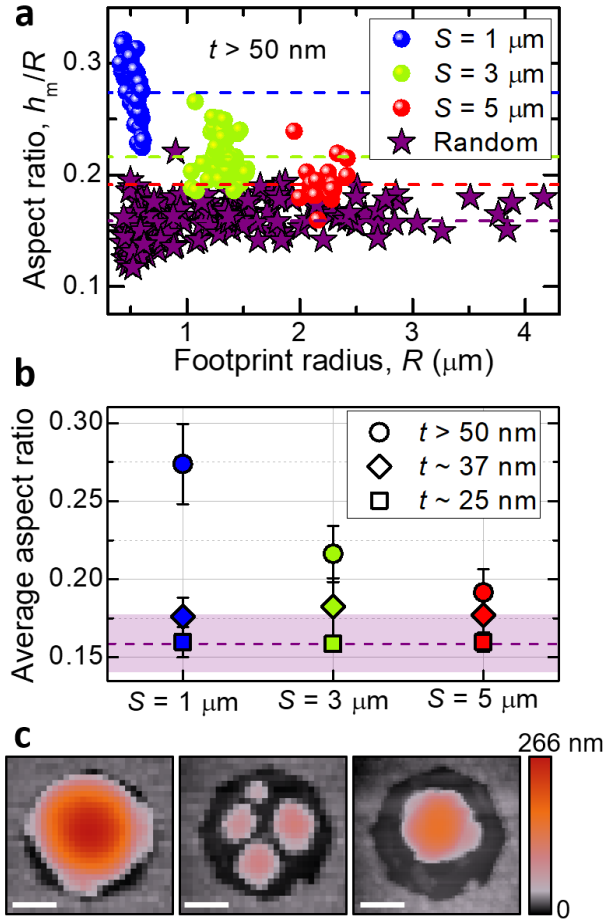

**Figure 4.** Increasing the aspect ratios and strains of the domes. (a) Aspect ratios measured by AFM in patterned MoS$_2$ domes (openings with size $S$ = 1 μm, blue dots; $S$ = 3 μm, green dots; and $S$ = 5 μm, red dots; resists with thickness 50 nm < $t$ < 100 nm), and on random MoS$_2$ domes (purple stars) as a function of the footprint radius. The coloured dashed lines highlight the average aspect ratios calculated for each set of data. (b) Average aspect ratio calculated by the data in panel a (circles) and in samples with thinner resists ($t$ = 37±6, rhombi; $t$ = 25±5, squares) for different sizes of the openings. The purple dashed line highlights the average aspect ratio measured for random domes, and the shaded area provides the associated uncertainty. (c) Left: AFM image of a patterned dome ($S$ = 3 μm) with $h_m/R$ = 0.206. Centre: AFM image acquired after indentation of the dome with the AFM tip. The dome was caused to deflate giving rise to four smaller domes with aspect ratio of about 0.16. Right: The four domes were induced to coalesce by the AFM tip, resulting in a single dome with $h_m/R$ = 0.169. The whole process took ~ 10 minutes. The white scalebar is 1 μm. The z range varies accordingly to the colorbar.

The general validity of the results of Fig. 3 is confirmed by the AFM statistical analysis of the aspect ratios shown in Fig. 4a, performed on several flakes patterned with masks of thickness $t$ between 50 and 100 nm, and on different samples. As a matter of fact, we obtain aspect ratios that range -on average- from 0.191±0.015 for domes in openings with $S$ = 5 μm, to 0.216±0.018 for domes where $S$ = 3 μm, to 0.274±0.026 for domes where $S$ = 1 μm, while in random domes we get 0.159±0.018. The average aspect ratio in patterned domes decreases to ~0.18 (almost independently of $S$) for $t$ ~37 nm, to approach the universal value for $t$ ~25 nm (see Figure 4b). Indeed, this effect is attributable to the mechanical constraint applied by the mask on the domes: If the mask is too thin, its contribution to the total energy of the system is negligible, so that the highly pressurized gas within the domes is able to raise the resist, as shown in Supplementary Figure S3; If the mask it thick enough, its contribution is instead significant and the aspect ratio of the domes is increased. As a further proof of this statement, we used the AFM tip to apply an extra mechanical stress to a patterned dome ($S$ = 3 μm) with $h_m/R$ = 0.206 (see Figure 4c, left). By pushing down the tip as described in Supplementary Figure 3, we induced a deflating process and separated the original dome in four smaller domes (Figure 4c, centre), all with an aspect ratio of about 0.16. Finally, by scanning the tip over the domes we were able to induce a coalescence phenomenon, resulting in the formation of a single dome (Figure 4c, right). This dome has smaller dimensions than the original one due to the initial deflation, and $h_m/R$ = 0.169 (see Supplementary Figure S4), which is very close the universal ratio.

Indeed, the high aspect ratios obtained in thick masks lead to high strains, which should result in significant shifts of the Raman modes[10,11,24,33,34] and exciton energies[8,9,10,11,13,17,18,24]. Therefore, micro-Raman (μ-Raman) measurements (see Experimental Section) were here used to confirm the presence of high isotropic biaxial strains acting at the summit of domes patterned in thick resists, like those in Figure 5a. In Figure 5b we show the average total in-plane strain at the summit of such domes ($\varepsilon_p^{summit}$, calculated via Hencky's model or, equivalently, by FEM calculations). While the structures created in openings with $S$ = 1 μm have the largest aspect ratio -with $\varepsilon_p^{summit}$ on average equal to (10.8±2.0) %-, the spatial resolution of our optical setup (our laser spot can be modelled as a gaussian with σ = 0.23 μm, see Experimental Section) is not enough to perform spatially-resolved optical measurements, as clear from Figure 5a, where the white scalebar corresponds to the size of the laser-spot. For this reason, domes in openings with $S$ = 3 μm were chosen, as the best trade-off between high ratios and acceptable spatial resolution.

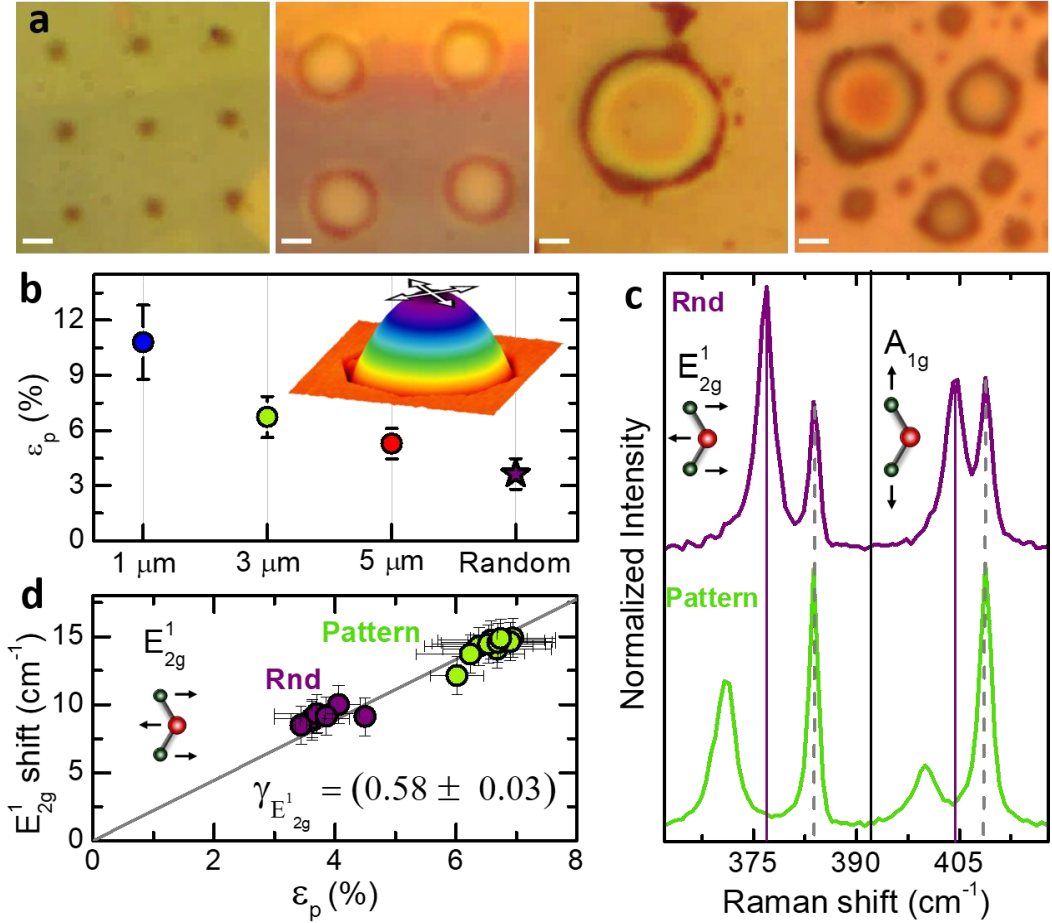

**Figure 5.** Micro-Raman measurements at the summit of random and patterned domes. (a) Optical images acquired with a 100× objective with NA=0.9 of (from left to right) patterned domes for $S = 1$ μm, for $S = 3$ μm, for $S = 5$ μm, and of random domes. The white scalebar is equal to 1.15 μm, that approximately corresponds to the size of our laser spot. (b) Average in-plane strain, $\varepsilon_p$, at the summit of patterned ($t > 50$ nm) and random domes, where strain is isotropic biaxial and reaches its maximum value. $\varepsilon_p$ is given by the sum of the two equivalent radial and circumferential components ($\varepsilon_r = \varepsilon_t$) by the Hencky's model, starting from the average values found for the aspect ratios in Figure 4a. Inset: AFM 3D image of a patterned dome depicting the two in-plane strain components (white arrows) at the dome summit. (c) Raman spectra acquired at the summit of a patterned ($S = 3$ μm) dome (bottom, green) and of a random dome with similar radius (top, purple) in the region of the in-plane $E_{2g}^1$ and out-of-plane $A_{1g}$ modes. A sketch of the two vibrational modes is displayed as inset. The grey dashed lines highlight the position of the peaks originating from the bulk flake, while the peaks at lower energy with respect to the bulk modes originate from the domes. The purple lines are set in correspondence of the modes of the random dome, to highlight the redshift of the modes of the patterned dome. (d) Redshifts for the in-plane mode (see sketch as inset) with respect to an unstrained ML measured on top of several random (purple) and patterned (green, $S = 3$ μm) domes with analogous dimensions, plotted as a function of the estimated in-plane strain. The grey line is a linear fit to the data, leading to the Grüneisen parameter $\gamma_{E_{2g}^1}$ reported as inset.

In order to have a one-to-one comparison between the optical measurements in random and patterned domes, Raman spectra were acquired at the maximum symmetry point for the domes - *i.e.*, their summit- as in Ref. 10. In Figure 5c we show a comparison between the μ-Raman spectrum acquired on top of a random dome (purple) and of a patterned dome (green), in the region of the in-plane $E_{2g}^1$ and out-of-plane $A_{1g}$ modes (see inset). The chosen random dome has diameter similar to the patterned dome. For each mode, the peaks at higher frequency -highlighted by dashed grey lines- originate from the bulk flake beneath the domes, while the lower-frequency peaks originate from the domes. Notably, Figure 5c points to a significant redshift of both the Raman modes in the patterned dome. To verify the consistency of our results with the measured aspect ratios, in Fig. 5d we plot the Raman redshifts for the $E_{2g}^1$ mode (with respect to a strain-free ML) as a function of $\varepsilon_p^{summit}$, measured for several patterned domes ($S = 3$ μm) and random domes with similar dimensions. As expected, the tensile biaxial strain induces a softening of the modes that varies linearly with strain, with a redshift rate of (2.2±0.1) cm$^{-1}$/% and a Grüneisen parameter[35] $\gamma_{E_{2g}^1} = 0.58\pm0.03$. This result agrees well with previous estimations of $\gamma_{E_{2g}^1} = 0.6$ in Ref. 33, 0.65 in Ref. 34, and 0.68 in Ref. 8, thus providing further confirmation of the enormous strains achieved with our method. The behaviour of the $A_{1g}$ mode, which is more subjected to strain-induced intensity modulations, is discussed in the Supplementary Information, Note 3. Finally, we also performed comparative micro-photoluminescence measurements on the domes, that also point out a much larger strain in patterned domes, as shown in Supplementary Figure S5.

**Conclusions.**

   In conclusion, we have demonstrated the possibility to create periodic configurations of giant, non-uniform strains in MoS$_2$. This result can be achieved by engineering the spontaneous process leading to the formation of 1-layer-thick domes in proton-irradiated bulk flakes, via deposition of polymeric masks and EBL patterning. Such a strategy allows, in the first instance, to gain control over the position and size of the domes, and can be brought to a further level of thoroughness ensuing the ability to design and realize the patterns properly. In particular, we have determined the irradiation conditions and properties of the masks that allow to mechanically-constrain the domes and enhance their aspect ratio remarkably. With this strategy, we have demonstrated the possibility to enhance the natural built-in strain of the domes in excess of a factor of 3, thus creating periodic high strain gradients in the MoS$_2$ membrane over micrometric regions. In light of the high strains achieved, this system might represent a unique platform for the observation of novel phenomena, enabling the characterization of the optoelectronic, valley- and spin-tronic fundamental properties of the $k$-space direct and indirect excitons in MoS$_2$ membranes under strongly anisotropic or perfectly isotropic strains, creating the conditions for ling-lived excitons[13] promising for Bose condensates[36], or engendering pseudo-magnetic fields[26,37,38] that could be potentially exploited for the generation of quasi-persistent currents[27]. Furthermore, the method proposed in this work is relevant for applications in photovoltaics, photocatalysis and photodetection[18], allowing to create the conditions for broadband absorption and harvesting of the photoexcited charge carriers by funnelling[13]. Finally, the technique presented in this letter is highly versatile, and might be exploited in diverse systems - possibly including other TMDs[24], h-BN[25], graphene[39] and much more besides- allowing the strain engineering of a wider class of two-dimensional materials.

**Experimental Section.**

1) *Electron-beam lithography patterning*:

The fabrication of H-opaque masks was performed by means of electron-beam lithography employing a Vistec EPBG 5HR system working at 100 kV. A hydrogen silesquioxane (HSQ) negative-tone e-beam resist was employed because of its property to be H-opaque under the irradiation conditions used in this work. Ordered arrays of octagonal openings with the desired diameter were patterned on a HSQ masking layer deposited on top of the sample surface. The thickness of the resist layer was controlled by varying the resist concentration and spinning speed. An electron dose of 150 $\mu C/cm^2$ and an aqueous development solution of tetramethyl ammonium hydroxide at 2.4% were used for the patterning of the HSQ masks.

2) *Atomic force microscopy measurements*:

AFM measurements were performed with two different instruments and by using different tips. The data were then analysed by different co-authors and led to analogous results. In particular, AFM measurements were performed by using:

- A Veeco Digital Instruments Dimension D3100 microscope equipped with a Nanoscope IIIa controller, employing Tapping Mode monolithic silicon probes with a nominal tip curvature radius of 5-10 nm and a force constant of 40 N/m;
- A Nanowizard III from JPK, equipped with Vortex controller, employing Tapping Mode silicon probes with a measured tip curvature radius of 50+/-10nm and a force constant of 40N/m.

The convolution of the tip used during the measurements was duly taken into account. All the data were taken at room temperature and under the same ambient conditions (atmospheric pressure). All the data were analysed with the Gwyddion software.

3) *Optical measurements*:

For second harmonic generation measurements, we used a supercontinuum laser tuned at ~900 nm with a ~50 ps pulse width and a 77.8 MHz repetition rate. The second harmonic signal was collected by means of a 750-mm focal length monochromator ACTON SP750 equipped with a 1200 groove/mm grating and detected by a back-illuminated Si CCD Camera (model 100BRX) by Princeton Instruments. A 100× objective with NA=0.9 was employed to excite and collect the light.

For Raman measurements the excitation laser was provided by a single frequency Nd:YVO4 lasers (DPSS series by Lasos) with emission wavelength equal to 532.2 nm. The Raman signal was spectrally analysed with the same monochromator and CCD described above. The micro-Raman (μ-Raman) spectral resolution was 0.7 cm$^{-1}$. The laser light was filtered out by a very sharp high-pass Razor edge filter at 535 nm (Semrock). The same objective described above was used for laser excitation/collection. The laser spot size was experimentally determined as follows: The laser was scanned across a reference sample, lithographically patterned with features of known width (1 μm). The intensity of the reflected light was fitted with the ideal reflectance profile, convolved with a Gaussian peak. The standard deviation of this peak, obtained as a fitting parameter, provides our estimate of σ = 0.23±0.01 μm.

For micro-photoluminescence (μ-PL) measurements the excitation laser was provided by a diode laser at 405 nm. Due to the poor efficiency of the emitted signal, the PL emission was spectrally analysed by means of a 200-mm focal length monochromator Isoplane160 equipped with a 150 groove/mm grating and detected by a back-illuminated Si CCD Camera (model 100BRX) by Princeton Instruments. The same objective described above was used for excitation/collection.

# 1. Acknowledgements.

We acknowledge support by Sapienza Università di Roma under the grants "Ricerche Ateneo" 2018 (A.P. and M.F.). M.F. and G.P. acknowledge support and funding from the Italian Ministry for Education, University and Research within the Futuro in Ricerca (FIRB) program (project DeLIGHTeD, Prot. RBFR12RS1W). E.B., A.P. and M.F. acknowledge funding from the Regione Lazio programme "Progetti di Gruppi di ricerca" legge Regionale n. 13/2008 (SINFONIA project, prot. n. 85-2017-15200) via LazioInnova spa. We acknowledge fund support from Australian Research Council (ARC) (numbers DE140100805 and DP180103238), and ARC Centre of Excellence in Future Low-Energy Electronics Technologies (project number CE170100039).

# Supplementary Information for

# Engineered creation of periodic giant, non-uniform strains in MoS₂ monolayers

Elena Blundo[†], Cinzia Di Giorgio[‡,§], Giorgio Pettinari[#], Tanju Yildirim[◊], Marco Felici[†], Yuerui Lu[◊], Fabrizio Bobba[‡,§] and Antonio Polimeni[†,*]

[†]Physics Department, Sapienza University of Rome, 00185 Rome, Italy.
E-mail: antonio.polimeni@roma1.infn.it

[‡]Department of Physics E.R: Caianiello, University of Salerno, 84084 Fisciano, Italy.

[§] SuPerconducting and other INnovative materials and devices institute, National Research Council, 84084 Fisciano, Italy.

[#]Institute for Photonics and Nanotechnologies, National Research Council, 00156 Rome, Italy.

[◊]Research School of Electrical, Energy and Materials Engineering, College of Engineering and Computer Science, The Australian National University, Canberra, ACT2601, Australia.

KEYWORDS: strain, MoS2, two-dimensional materials, engineering, domes, patterning.

**Contents:**

**Supplementary Note 1: Engineering the patterning procedure**

In panel A of Figure 1 of this Note we show the sketch of a typical pattern designed for our samples. The pattern is realized by depositing a film with uniform thickness of hydrogen silsesquioxane (HSQ) on the sample. A negative tone e-beam resist is used, so that the desired pattern (wine areas in the figure) can be obtained by electron-irradiating the whole sample's surface apart for the white areas visible in the figure (octagonal openings and corridors) where the domes will form upon proton-irradiation.

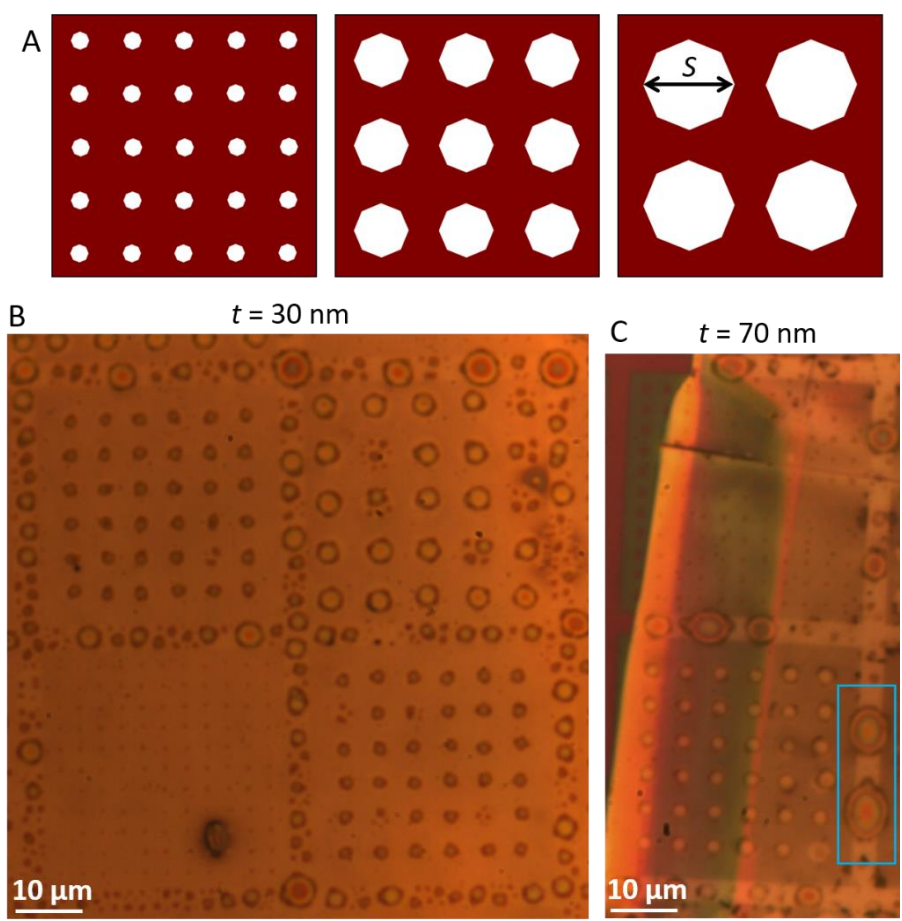

**Figure 1 of Note 1. Design and realization of the HSQ masks.**
**(A)** Sketch of a typical mask designed for patterning the samples: arrays of openings (in white) with different characteristics are arranged in different squares (in wine) with same dimensions. $S$ indicates the diameter of the openings. The sketch here shown might represent a basic module to be repeated periodically all over the sample. **(B)** Optical image of a $MoS_2$ sample after patterning (where the thickness of the mask, $t$, is about 30 nm) and dome formation. Here the opening sizes in the four squares are 3 μm (on a 6 x 6 array, top left), 5 μm (on a 5 x 5 array, top right), 1 μm (on a 11 x 11 array, bottom left) and 3 μm (on a 6 x 6 array, bottom right). Indeed, many domes with random sizes formed within the corridors that separate the squares. **(C)** Optical image of another $MoS_2$ sample after patterning (with a 70-nm-thick mask) and dome formation. Here the opening sizes in the two squares are 1 μm (on a 11 x 11 array, top) and 3 μm (on a 6 x 6 array, bottom). In this case, many domes in the corridors (see cyan rectangle) are elliptical due to the anisotropic constraint due to the thick mask around.

The choice of realizing octagonal openings rather than circular ones is aimed at reducing the electron beam lithography exposure time while still creating openings with a circular-like shape. After the electron exposure, the resist is developed in an aqueous solution of tetramethyl ammonium hydroxide at 2.4% to remove the unexposed resist and leave the desired openings (see Experimental Section in the main text). With this procedure, several arrays with openings (white areas) of different sizes are arranged in different squares (in wine) with same dimensions. The size of the openings, their number and disposition, and the distance between the squares can be varied according to the specific needs. Once the basic ingredients are chosen, the same module (like the one represented in panel A) is repeated periodically all over the sample surface where the flakes had been previously deposited.

An optical image of a patterned $MoS_2$ sample (with a 30-nm-thick mask) -after dome formation- is shown in panel B. Here the size of the squares, and thus the number of openings, is larger than in the sketch of panel A. The domes formed within the openings have typically a regular size, while the presence of domes with random size can be observed in the corridors between the squares. In panel C, we show optical image of another patterned $MoS_2$ sample. In this case the thickness of the mask is about 70 nm. The domes within the openings are almost totally filling the openings. In this case, the thick mask represents a constraint for the domes; this is particularly is evident by looking at the corridors between the different squares, where elliptically-deformed domes formed, such as those withing the cyan rectangle.

As for the thickness of the resist, masks with different thicknesses were realized. While the height of the mask is not perfectly constant all over the sample –due to the small sizes the of samples that are typically processed and to the presence of the flakes on the surface– with our method we are able to realize masks of a certain height within about 10 nm. In any event, the heights of the masks were all measured by AFM, flake by flake, so that the thickness values quoted in this work were all determined experimentally. Our studies show that for masks with height between 50 and 100 nm, we can achieve high aspect ratios of the domes, still having a dome formation process characterized by a high yield, > 95 % for medium and small domes and ~ 60-70 % for big domes, as one can see in panels B and C. The yield of the formation process is indeed also influenced by the proton-dose. For low doses, the domes form almost within all the openings. If the dose is brought to an optimum level, the domes almost totally fill the openings, as shown in panel C and in Figure 1 of the main text. However, if higher doses are used, the domes start exploding, as a consequence of the too high internal pressure exerted on the membrane by the trapped hydrogen, and of the boundary role played by the mask.

**Supplementary Note 2: Patterned domes and strain fields.**

As discussed in Refs. 1 and 2, finite element method (FEM) calculations can be used to model the height profile and strain tensor -in spherical coordinates- of the domes. This allows us to estimate how the strain tensor varies across the dome surface. In Figure 1A of this Note (upper panel) we show the simulated strain tensor components as a function of the distance from the dome centre, for the same random dome of Fig. 3 of the main text (with footprint radius $R$ = 2200 nm and maximum height $h_m$ = 363 nm, resulting in $h_m/R$ = 0.165). In the lower panel we calculate the degree of isotropy of the in-plane strain acting on the dome's surface, defined as:

$$\text{Strain Isotropy} = 1 - \frac{\varepsilon_r - \varepsilon_t}{\varepsilon_r + \varepsilon_t}.$$

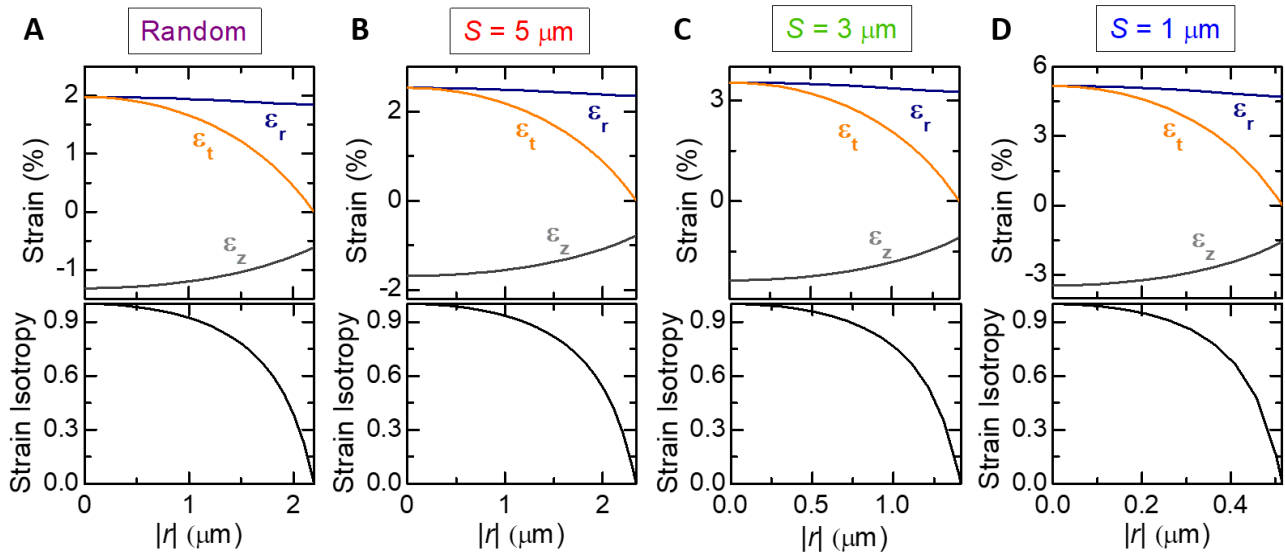

**Figure 1 of Note 2. Strain distribution in patterned domes.**
**(A) Upper panel:** Strain components in polar coordinates ($\varepsilon_r$ = radial component, $\varepsilon_t$ = circumferential component, $\varepsilon_z$ = perpendicular component[1]) calculated for the same random dome of Fig. 3 of the main text. The components were calculated via FEM calculations as a function of the distance from the centre of the dome, $|r|$. **Lower panel:** Degree of isotropy of the in-plane strain acting on the dome. **(B-D)** Same for the three patterned domes of Fig. 3 of the main text. The three domes have: $R$ = 2343 nm and $h_m$ = 436 nm, resulting in $h_m/R$ = 0.186 (B); $R$ = 1414 nm and $h_m$ = 310 nm, resulting in $h_m/R$ = 0.219 (C); $R$ = 515.5 nm and $h_m$ = 136.1 nm, resulting in $h_m/R$ = 0.264 (D).

While the sum of the two in-plane components (*i.e.*, the radial, $\varepsilon_r$, and circumferential, $\varepsilon_t$, components[1]) is the relevant quantity to be considered for the strain-induced variation of the electronic properties of TMDs, the degree of isotropy should be taken into account when considering the vibrational properties of the strained membrane and its intertwined spin/valley degrees of freedom. As shown here, according to our calculations strain is isotropic at the center of the domes, while it becomes uniaxial -and therefore totally anisotropic- at the edges. The real behavior at the edges might not be perfectly described by our model, due to both the transition from bulk to dome, the boundary role played by the mask in patterned domes, and the presence of small domes at the edges of the large domes (typically observed mostly in random domes), that might lead to deviations from the ideal situations described by numerical simulations[2]. In any event, one should expect strain to be strongly anisotropic at the edges. The same calculations of panel A are repeated for the three patterned domes of Fig. 3 of the main text in panels b-d, showing that the same qualitative behaviour holds for both the patterned and the random domes. Quantitively, there is instead a remarkable difference, with very high strain values achieved for small patterned domes, close to the rupture critical value, according to previous studies[3,4]. Such a result is of great interest for both fundamental and applicative prospects. The creation of high and strongly anisotropic strains can give rise to exciton anisotropy splittings of interest for valley- and spin-tronics[5]. The possibility to generate strain gradients over micrometric region is potentially interesting for applications in the fields of photovoltaics, photocatalysis and photodetection[6]. In particular, strain has been demonstrated to change the band structure of TMDs in such a way that the exciton energy shifts by several tens of meV/%[2,4,7]. Additionally, in presence of a seamless strain variation, the band gap reduction for increasing strain induces funnelling phenomena, where the excitons drift for hundreds of nm before recombining[2,6,8]. In addition, it has been recently demonstrated that strains between 1 and 4 % in $WS_2$, $MoS_2$ and $WSe_2$ turn the band gap from direct to indirect, with indirect excitons characterized by much longer decay times[2]. The generation of periodic and high strain gradients allow therefore at the same time to absorb energy over a relatively large portion of the electromagnetic spectrum and to harvest the photogenerated carriers in particular position, which is particularly promising for the realization of efficient solar cells[6].

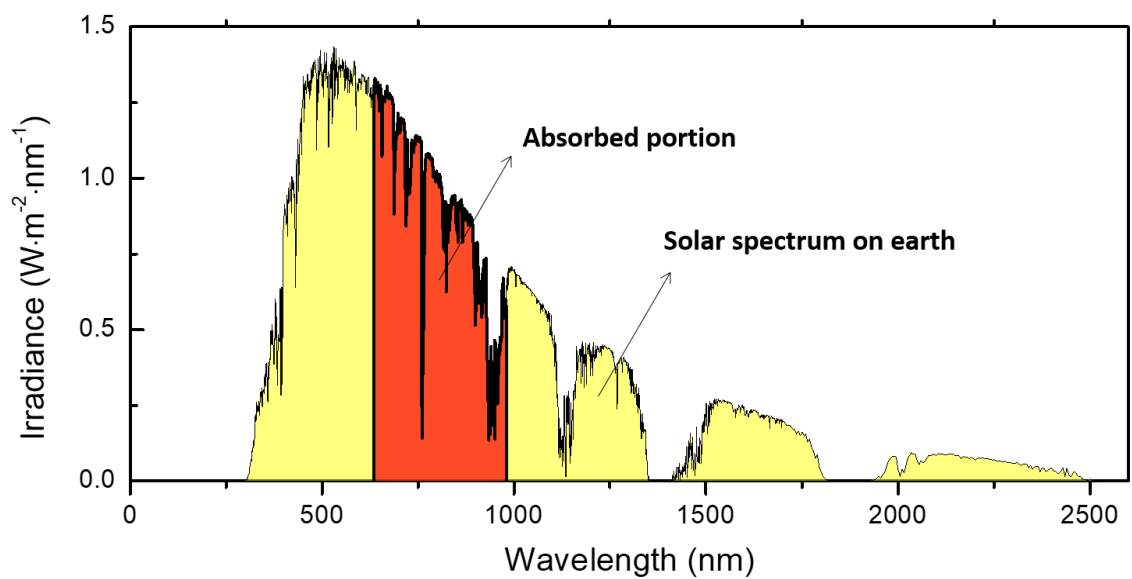

**Figure 2 of Note 2. Absorption of a wide portion of the solar spectrum on earth by patterned domes.** The yellow area represents the solar spectrum on earth according to the active standard ASTM G173 - 03(2012). The red area represents approximately the portion of the spectrum that would be absorbed by medium patterned domes, according to the shift rates estimated for the direct and indirect excitons in Ref. 2.

As for the possibility to absorb light, with patterned arrays of medium domes, for instance, we get a variable total in-plane strain between 3 and 7 %. According to the shift rates for the excitons measured in Ref. 2, this would allow to absorb energy in a wide wavelength range, thus covering a significant portion of the solar spectrum, as shown in Figure 2 of this Note.

In addition to the great potentiality of our system for photovoltaics applications, the possibility to generate strain fields in controllable positions holds great relevance also because it paves the way for the observation of novel physical phenomena. In particular, strain fields in graphene have been predicted and demonstrated to give rise to pseudomagnetic fields, up to hundreds of Tesla[9,10,11,12]. Additionally, periodic configurations of magnetic fields in 2D electron gases have been predicted to allow current to flow without dissipation[13], which is similar to superconductivity. While the insurgence of pseudomagnetic fields in strained graphene has been assessed as a manifestation of the chirality of the Bloch wave functions around the two inequivalent corners of the hexagonal Brillouin zone, TMD MLs combine these effects with a strong spin-orbit coupling (SOC)[14]. As a consequence, it has been predicted that in mechanically deformed TMD MLs the large spin-orbit coupling rotates the wave function in the spinor basis, and a gauge field may arise[15].

**Supplementary Note 3: Analysis of the Raman shifts in random and patterned domes.**

Here we discuss micro-Raman (μ-Raman) measurements aimed at confirming that our domes are subjected to high strains as a consequence of the high aspect ratios measured by AFM. As a general fact, strain has been demonstrated to lead to a softening of the Raman modes[1,2,4,7,16,17,18]. The quantitative effect of strain on the Raman modes, however, is dependent on the kind of strain that is applied (i.e., its orientation and anisotropy). For this reason, we performed our measurements in the point of maximum symmetry for our domes, that is, the centre. There, strain is known to be isotropic biaxial and a simple correspondence with the aspect ratio has been established by models developed in the framework of the membrane theory[1,2,4,19,20]: $\varepsilon_p = 2 \cdot f(\nu) \cdot (h_m/R)^2$, with $f(\nu) = 0.721$ for MoS$_2$ domes, where $\varepsilon_p$ is the total in-plane strain. As for the patterned domes, in the case of the small domes (opening with size $S = 1$ μm) we were limited by the resolution of our optical setup (the laser spot has dimensions comparable to the opening, see Experimental Section in the main text and Figure 5a of the man text). For big domes, we would have had the best resolution, but the aspect ratio is only marginally larger than in random domes. For this reason, medium domes in openings with $S = 3$ μm represented the best trade-off, having significantly higher aspect ratios and allowing to be measured with relatively good spatial resolution. As for random domes, in order to have a comparison on an equal footing, domes with diameter of about 3 μm were chosen. The results obtained for the in-plane E$_{2g}^1$ mode have been discussed in the main text. The redshifts measured for the out-of-plane A$_{1g}$ Raman peak (see, e.g., spectra in Fig. 5c of the main text) with respect to the strain-free ML at the summit of random domes (purple) and patterned domes (green) are instead displayed in Figure 1 of this Note.

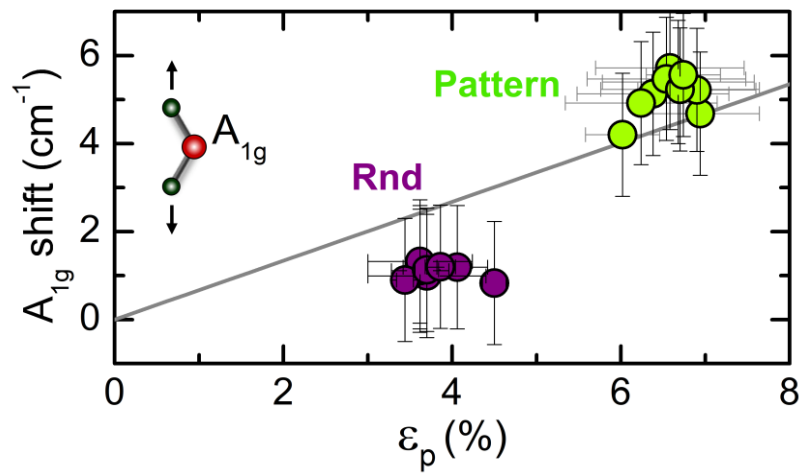

**Figure 1 of Note 3. Redshift of the A$_{1g}$ Raman peak in random and patterned domes.**
Redshifts measured for the A$_{1g}$ Raman peak (see, e.g., spectra in Fig. 5c of the main text) with respect to the strain-free ML at the summit of patterned (openings with size $S = 3$ μm) domes (green) and of random domes with similar dimensions (purple). The data are plotted as a function of the total in-plain strain at the summit, $\varepsilon_p^{summit}$, analogously to Fig. 5d of the main text. The grey line is a linear fit to the data.

The grey line is a linear fit to the data, showing that the data do not follow the expected linear behaviour with strain. A deviation from linearity has been observed also in WS$_2$ domes in Ref 16, and can be ascribed to the strong intensity modulation of this mode with strain, as shown and discussed in Refs. 4 and 16. In particular, notice that according to Ref. 4 the intensity of the A$_{1g}$ mode exhibits a maximum for a biaxial strain ~ 3 % (corresponding to $\varepsilon_p^{summit}$ ~ 6 %). and a minimum for a biaxial strain ~ 2 % (corresponding to $\varepsilon_p^{summit}$ ~ 4 %)., that approximately correspond to the strain values at the top of patterned and random domes, respectively. This effect, coupled to the non-optimum resolution of our setup compared to the dimensions of the domes, might cause the data concerning random domes to deviate from the expected linear behaviour.

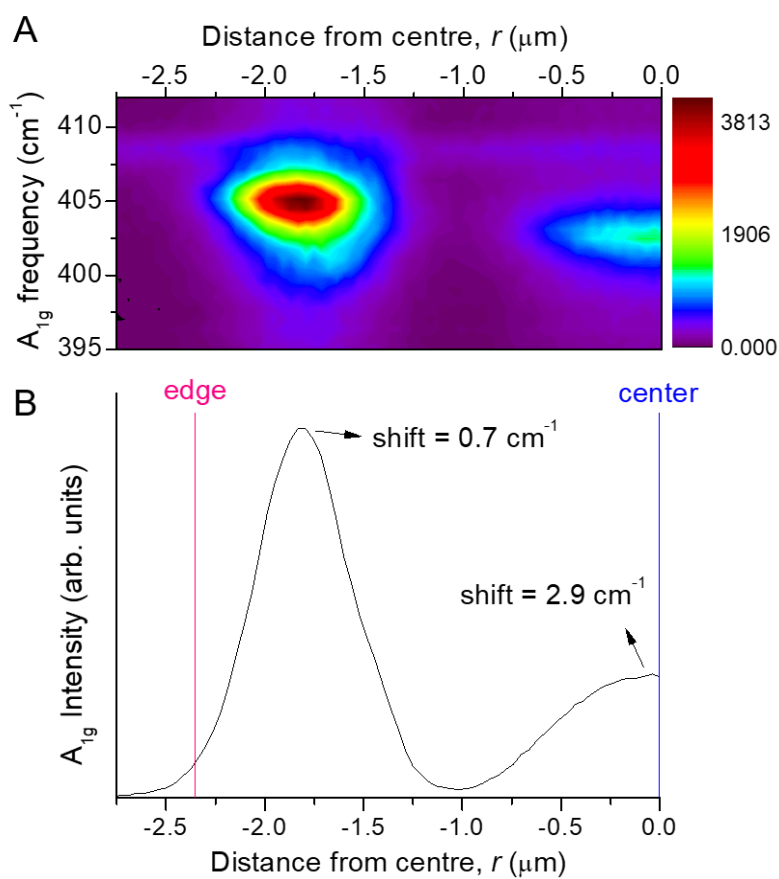

**Figure 2 of Note 3. Raman mapping along a radius of a large random dome.**
**(A)** Colormap corresponding to a μ-Raman scan across a random MoS$_2$ dome while moving from an edge towards the centre. A relatively large dome with radius 2.35 μm was chosen in order to achieve a good spatial resolution. The horizontal axis indicates the laser spot position with respect to the dome centre (*r*), whereas the vertical axis indicates the Raman shift with respect to the laser line (laser wavelength = 532.2 nm, see Experimental Section in the main text). The μ-Raman intensity is shown in a false colour scale (see colour bar). The mode at 408.6 cm$^{-1}$ is the A$_{1g}$ mode of the bulk flake beneath the dome. The A$_{1g}$ mode of the dome clearly features both a redshift, while moving from the edge towards the centre, and a strong modulation of intensity, which is ascribable to the strain variation along the scanning direction[4]. **(B)** Integrated intensity of the A$_{1g}$ mode for the dome shown in panel B. The pink and blue lines highlight the positions of the edge and the centre of the dome, respectively.

To verify this hypothesis, we acquired the Raman spectra at the summit of random domes with larger diameter, that allows to get better resolved measurements. In Figure 2A of this Note we show a μ-Raman scan across a random $MoS_2$ dome while moving from an edge towards the centre, for a dome with radius 2.35 μm and $h_m/R = 0.164$. The faint horizontal trace at 408.6 $cm^{-1}$ is the $A_{1g}$ mode of the bulk flake beneath the dome. This mode features a tiny variation of intensity, ascribable to interferential phenomena, and remains constant in energy. The $A_{1g}$ mode of the dome, on the contrary, clearly features both a redshift (while moving from the edge towards the centre) and a strong modulation of intensity, which is therefore chiefly ascribable to the strain variation along the scanning direction[4]. To better appreciate the intensity variation while going from the edge to the centre, in panel B we display the integrated intensity as a function of the radial coordinate, $r$. Indeed, the maximum intensity is reached close to the edge, where the shift with respect to the unstrained ML is 0.7 $cm^{-1}$. The intensity drops at a distance of about 1 μm from the centre, to re-increase a little bit towards the centre, where the Raman shift equals 2.9 $cm^{-1}$. Notice that for the smaller random domes in Figure 1 of this Note the shift measured at their summit is also equal to 0.7 $cm^{-1}$. This value is the same found towards the edge of the larger dome, where the intensity is maximum. As a matter of fact, when measuring smaller domes, the Raman spectrum is the result of a convolution within the exciting laser spot (gaussian with σ = 0.23 μm, see Experimental Section in the main text), and the component with maximum intensity could dominate. Furthermore, the relative intensity of the two maxima in Figure 2B could be different in smaller domes (due to the different role played by interference[16]) and further favour the edge component. This is possibly the reason for the deviation from the expected behaviour for this mode, and the different shift measured for random domes with analogous aspect ratio but different radii (shift ~ 0.7 $cm^{-1}$ in domes with diameter ~ 3 μm, as shown in Figure 1, and 2.9 $cm^{-1}$ in the dome with diameter ~ 4.7 μm discussed in Figure 2). The in-plane $E_{2g}^1$ mode could be as well affected by a similar phenomenology, but the intensity of this mode with strain has been shown to be less subjected to oscillating behaviours for the strain values considered[4].

In order to have a quantitative feedback of the above discussion, in Table 1 of this Note we report the redshift rates and Grüneisen parameters for the $E_{2g}^1$ and $A_{1g}$ Raman modes, calculated by considering only random domes, only patterned domes, and all the domes. For the reasons discussed above, a large difference in the redshift rate -and consequently in the Grüneisen parameter- is found for the $A_{1g}$ mode by considering the different sets of data, where the data concerning random domes are affected by the oscillating behaviour of this mode with strain. On the contrary, the redshift rate of the $E_{2g}^1$ mode varies little depending on the set of data considered, confirming that this mode is less subjected to oscillations in the intensity for increasing strain. Notice also that the values found for the $E_{2g}^1$ mode are in agreement with those reported in previous works, as discussed in the main text. In addition, the Grüneisen parameter found for the $A_{1g}$ mode in patterned domes agrees well with previous results: $\gamma_{A_{1g}}$ = 0.21 in Ref. 4 where strain is biaxial and $\gamma_{A_{1g}}$ = 0.21 in ref. 21 where strain is uniaxial.

|  | **Only random** | **Only patterned** | **All data** |
|---|---|---|---|
| $\Delta_{E_{2g}^1}$ (cm$^{-1}$/%) | 2.40±0.05 | 2.15±0.05 | 2.2±0.1 |
| $\gamma_{E_{2g}^1}$ | 0.62±0.02 | 0.56±0.02 | 0.58±0.03 |
| $\Delta_{A_{1g}}$ (cm$^{-1}$/%) | 0.30±0.05 | 0.80±0.05 | 0.65±0.10 |
| $\gamma_{A_{1g}}$ | 0.07±0.01 | 0.20±0.01 | 0.17±0.03 |

**Table 1 of Note 3. Redshift rates and Grüneisen parameters.**
Redshift rates for in-plane strain ($\Delta$) and Grüneisen parameters ($\gamma$) for the in-plane $E_{2g}^1$ and out-of-plane $A_{1g}$ Raman modes in MoS$_2$ monolayers under isotropic strain (domes' summit). The displayed values were calculated from the data in Fig. 5d of the main text and Figure 1 of this Note, by considering only the set of data concerning random domes, only that concerning patterned domes, and both the two sets.

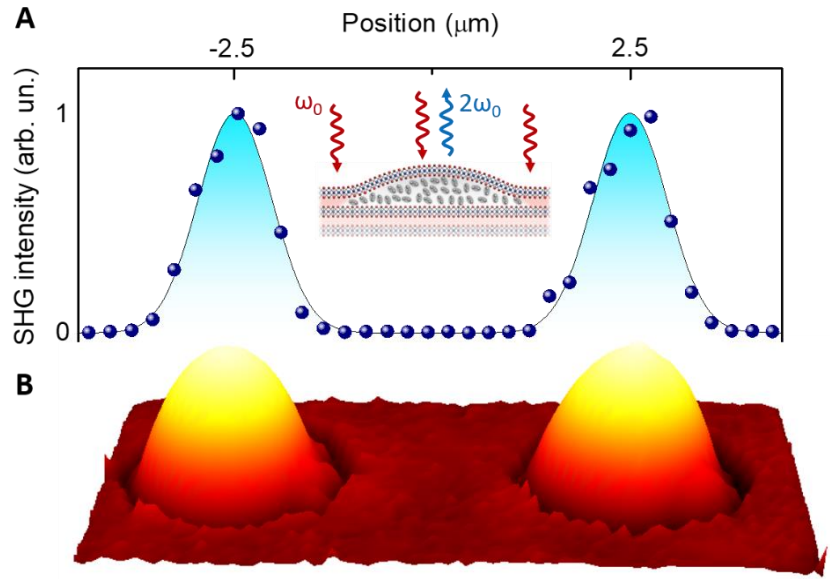

**Supplementary Figure S1: Probing the thickness of the domes via second harmonic generation.**

**(A)** Intensity of the second harmonic (SH) signal (navy points) during a scan performed on a patterned sample ($S$ = 3 μm). The measurements were performed by exciting at 900 nm and collecting the signal at 450 nm (see Experimental Section in the main text). The scan was performed by moving along a given direction at steps of 250 nm, in such a way to acquire the measurements along the diameter of two domes. The correspondence between the laser spot position and the domes is established by panel B. As here shown, second harmonic generation (SHG) is obtained only when scanning the laser over the two domes. The solid line is a fit to the intensity data with gaussian functions. Indeed, $MoS_2$ is characterized by the ability to originate bright SHG in the monolayer limit, while the SH signal decreases rapidly as the number of layers increases[22,23]. In particular, even number of layers do not give rise to SHG, while the SH of the tri-layer is about 5 times lower than that of the ML. SHG measurements can be therefore exploited to estimate the number of layers and were thus performed on several domes, as for the two domes shown here. The intensity of the SH signal was found to be always comparable or larger than in the ML for all the measured domes, thus suggesting a single-layer thickness of the domes, as schematized in the inset. **(B)** 3D AFM image of the two domes on which SHG measurements were performed.

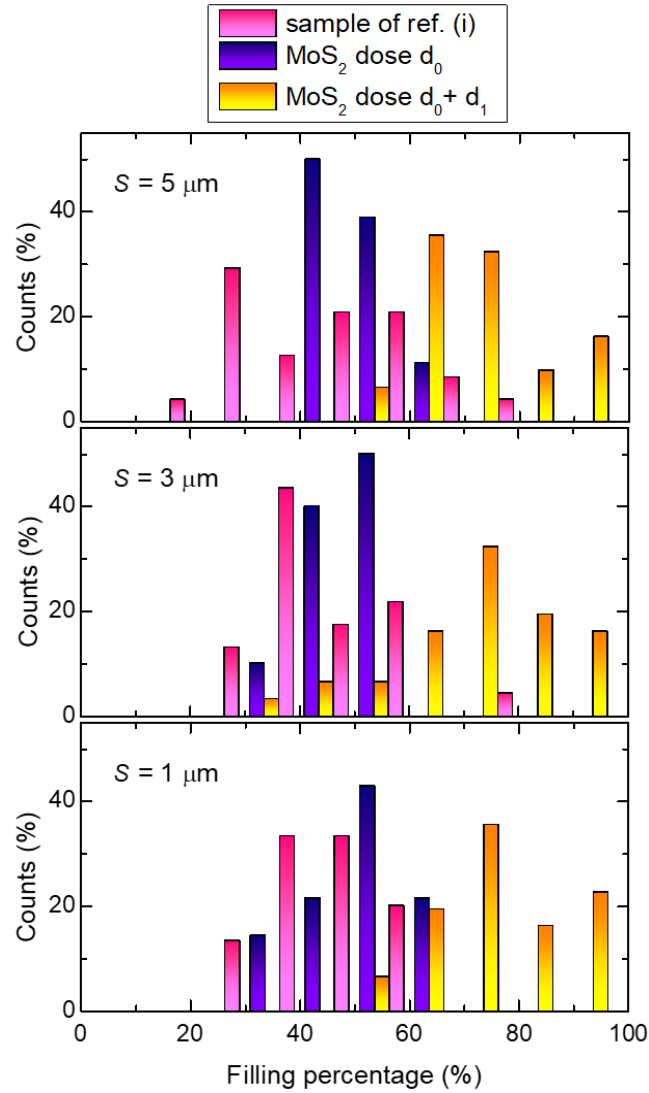

**Supplementary Figure S2: Filling percentages of the openings for increasing proton-doses.**

Filing percentages for patterned domes in openings with size $S = 5$ μm (upper panel), 3 μm (panel in the middle), and 1 μm (lower panel). The percentages were calculated by estimating the footprint of the domes (via AFM measurements of optical images for large domes) and by dividing it for the area of the opening. The filling percentages were calculated for: (i) The patterned $WS_2$ sample of ref. 1, which had been irradiated with dose $d_i = 4.0 \cdot 10^{16}$ protons/cm$^2$ (pink columns); (ii) A $MoS_2$ sample irradiated with dose $d_0 = 5.5 \cdot 10^{16}$ protons/cm$^2$ (dark-blue columns); (iii) the same $MoS_2$ sample, re-irradiated with an extra-dose $d_1 = 2.0 \cdot 10^{16}$ protons/cm$^2$ (orange columns). Indeed, the higher the dose the higher the filling percentages, on average. For each set of data, approximately 20 domes were considered.

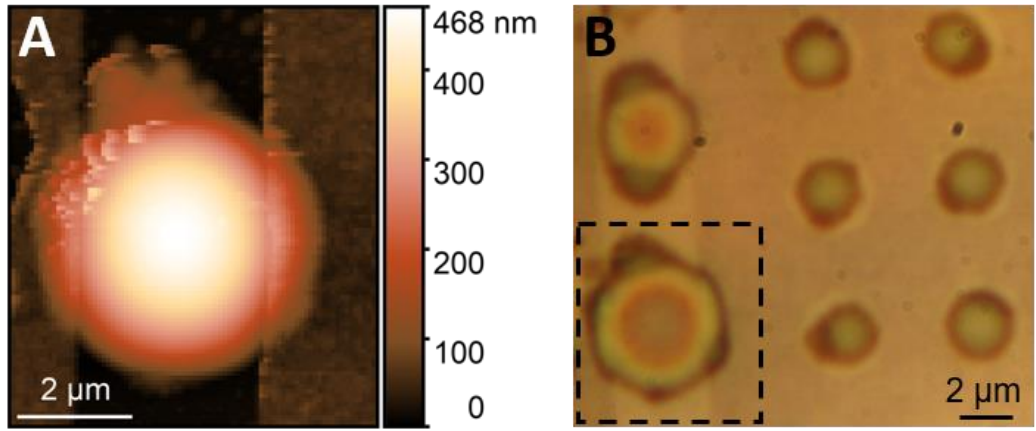

**Supplementary Figure S3: Raising of the thin masks.**

**(A)** AFM image of a dome created on a flake patterned with a 45-nm-thick mask. The dome formed within a corridor, and was able to raise the mask at its edges. A similar phenomenology is often observed in samples patterned with thin resists (below 50 nm) and irradiated with doses high enough to fill the openings. **(B)** Optical image of the same sample of panel A. The AFM image was acquired approximately in the area within the black dashed line.

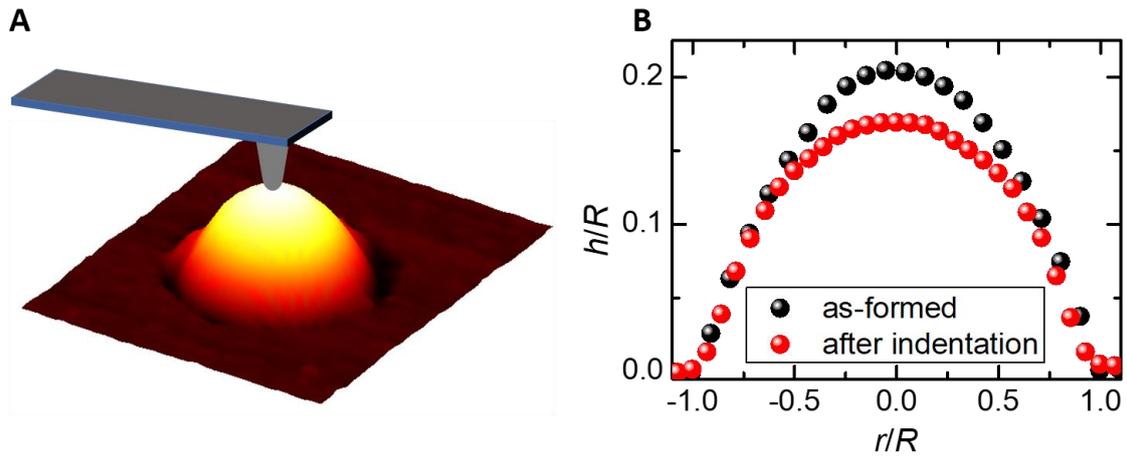

**Supplementary Figure S4: Indentation of a dome by the AFM tip.**

**(A)** Sketch of the indentation experiment by the AFM tip. A tip with curvature radius equal to 40 nm is positioned at the centre of the dome (same dome of Figure 4c, left, in the main text) and pushed down until reaching the substrate. **(B)** AFM profile along a diameter of the dome before indentation (black points) and after the indentation experiment (red points, corresponding to the AFM image of Figure 4c, right panel). The height profiles have been normalized to the footprint radii of the dome to highlight the different aspect ratios.

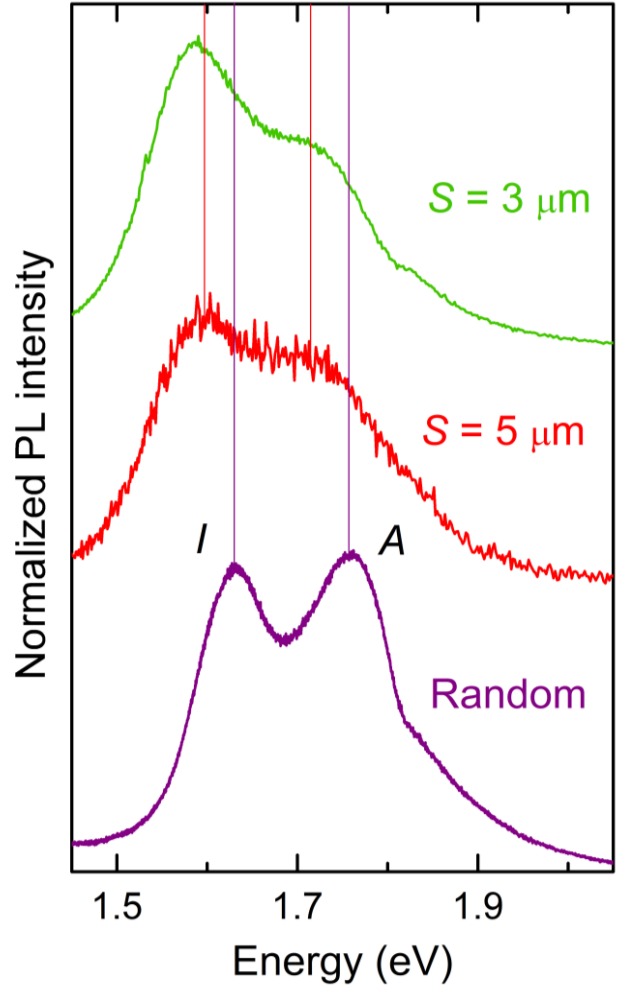

**Supplementary Figure S5: Micro-photoluminescence on patterned domes.**

Micro-photoluminescence (μ-PL, see Experimental Section in the main text) spectra acquired at the summit of patterned and random domes. The spectrum concerning the random dome (bottom, purple) is the same of Ref. 2 and is plotted here for comparison. This spectrum was acquired on a dome with $h_m/R$ = 0.164 (corresponding to $\varepsilon_p^{summit}$ = 3.9 %). While the unstrained MoS$_2$ ML is characterized by a direct gap (involving the K point of both conduction band, CB, and valence bad, VB), strains higher than ~1.8 % turn the band gap from direct to indirect[2] (the maximum of the VB turns from the K point to the Γ point[24]). Here, the strain is higher than the crossover value, and both the direct (*A*, corresponding to the K$_{CB}$-K$_{VB}$ transition) and indirect (*I*, corresponding to the K$_{CB}$-Γ$_{VB}$ transition) exciton transitions can be observed. The purple solid lines highlight the position of the indirect and direct excitons for the random dome. In the middle, we show the spectrum (in red) of a dome created within openings with size $S$ = 5 μm. The dome has $h_m/R$ = 0.180 ($\varepsilon_p^{summit}$ = 4.7 %), and an evident strain-induced redshift of both the exciton transitions can be observed. The red solid lines highlight the position of the exciton peaks, as a guideline for the eye. Finally, the green spectrum on top was acquired on a dome created in an opening with size $S$ = 3 μm, and is characterized by $h_m/R$ = 0.213 ($\varepsilon_p^{summit}$ = 6.5 %). In this case, the exciton transitions are further redshifted, with a lower shift-rate that is due to the non-optimum resolution of our optical setup compared to the dimension of the dome (see Fig. 4a in the main text and discussion in Supplementary Note 3).